\pdfoutput=1
\documentclass[a4paper,11pt]{article}
\usepackage{pos}
\usepackage{subcaption}

\usepackage[shortlabels]{enumitem}

\title{SUSY in the light of the new ``MUON G-2'' Result}

\author*[a]{Manimala Chakraborti}
\author[b]{Sven Heinemeyer}
\author[c]{Ipsita Saha}

\affiliation[a]{Astrocent, Nicolaus Copernicus Astronomical Center of the Polish                                                                                                                                       Academy of Sciences, \\ul.\ Rektorska 4, 00-614 Warsaw, Poland}

\affiliation[b]{Instituto de F\'isica Te\'orica (UAM/CSIC),
Universidad Aut\'onoma de Madrid, 
Cantoblanco, \\28049, Madrid, Spain}

\affiliation[c]{Kavli IPMU (WPI), UTIAS, University of Tokyo, Kashiwa, Chiba 277-8583, Japan}

\emailAdd{manimala@camk.edu.pl}
\emailAdd{Sven.Heinemeyer@cern.ch}
\emailAdd{ipsita.saha@ipmu.jp}

\abstract{The recently published result from the Fermilab ``MUON G-2''  experiment
has confirmed the persistent $3-4\,\sig$ discrepancy between the experimental result from BNL
for the anomalous magnetic moment of the
muon, \gmin2, and its Standard Model (SM) prediction.
The combination of the two measurements yields a deviation of
$\newdiffsig\,\sig$
from the SM value.
Here, we review the parameter space of the
electroweak (EW) sector of the Minimal Supersymmetric Standard Model
(MSSM), that can accommodate the
anomaly while being in full agreement with other experimental
data, particularly the
direct searches for EW particles at the LHC and dark matter (DM)
relic density and direct detection constraints.
We find that the combined constraints set an upper limit
of $\sim 600 \gev$ for
the LSP and NLSP masses establishing clear targets for
the future collider searches.
}

\FullConference{%
  *** The European Physical Society Conference on High Energy Physics (EPS-HEP2021), ***\\
  *** 26-30 July 2021 ***\\
  *** Online conference, jointly organized by Universität Hamburg and the research center DESY ***
}

\newcommand\tb{\tan\beta}

\newcommand\ReDiag{\mathop{%
  \raise .5pt\hbox{[}%
  \widetilde{\mathrm{Re}}%
  \raise .5pt\hbox{]}}}
\newcommand\ReOffDiag{\mathop{%
  \raise .5pt\hbox{$\llbracket$}%
  \widetilde{\mathrm{Re}}%
  \raise .5pt\hbox{$\rrbracket$}}}

\newcommand\Mh{M_h}

\newcommand\MA{M_A}

\newcommand\Sn{\tilde\nu}
\newcommand\Sl{\tilde l}

\newcommand\Slpm{\tilde l^\pm}

\newcommand\Sel[1]{\tilde e_{#1}}

\newcommand\msl[1]{m_{\Sl_{#1}}}

\newcommand\Stau[1]{{\tilde\tau_{#1}}}

\newcommand\mL{m_{\tilde l_L}}
\newcommand\mR{m_{\tilde l_R}}

\newcommand\ino[1]{\tilde\chi_{#1}}

\newcommand\chapm[1]{\ino{#1}^\pm}

\newcommand\cha{\chapm}
\newcommand\mcha[1]{m_{\chapm{#1}}}

\newcommand\neu[1]{\ino{#1}^0}
\newcommand\mneu[1]{m_{\neu{#1}}}

\newcommand\refeq[1]{Eq.~(\ref{#1})}

\newcommand\citere[1]{Ref.~\cite{#1}}
\newcommand\citeres[1]{Refs.~\cite{#1}}

\newcommand{\CP}{{\cal CP}}
\newcommand{\cp}{{\CP}}

\newcommand{\tev}{\,\, \mathrm{TeV}}
\newcommand{\gev}{\,\, \mathrm{GeV}}

\newcommand\CM{\texttt{CheckMATE}}

\newcommand\msmu[1]{m_{\tilde{\mu}_{#1}}}
\newcommand\mstau[1]{m_{\tilde{\tau}_{#1}}}

\newcommand{\sig}{\sigma}

\def\order#1{\ensuremath{{\cal O}(#1)}}
\def\reffi#1{\mbox{Fig.~\ref{#1}}}

\def\De{\Delta}

\def\gmin2{\ensuremath{(g-2)_\mu}}
\def\amu{\ensuremath{a_\mu}}

\newcommand{\ssi}{\ensuremath{\sig_p^{\rm SI}}}

\definecolor{Orange}{named}{orange}
\definecolor{Purple}{named}{purple}
\definecolor{Lightblue}{cmyk}{0.9,0.1,0.1,0.3}
\definecolor{dgelborange}{cmyk}{0.,0.3,0.5, 0.}
\definecolor{Lila}{rgb}{0.5,0.,1}
\definecolor{Darkgreen}{rgb}{0.,.7,0.2}

\graphicspath{{plots/}}

\newcommand{\fnalbnlsig}{\htrs{0.8}}   
\newcommand{\newdiff}{\htrs{25.1}}    
\newcommand{\newdiffunc}{\htrs{5.9}}   
\newcommand{\newdiffsig}{\htrs{4.2}}   
\newcommand{\htrs}[1]{{\color{black} #1}}

\begin{document}
\maketitle

\section{Introduction}
\label{sec:intro}
The sustained deviation of $3$-$4\,\sigma$ in the anomalous magnetic
moment of muon, \gmin2, 
between the theoretical prediction of the SM~\cite{Aoyama:2020ynm}
(see \citere{CHS3} for a full list of references) and
the experimental observation by the Brookhaven National Laboratory
(BNL)~\cite{Bennett:2006fi} has long been hinting towards the existence of some new
physics scenario.
The new result from the Fermilab
``MUON G-2'' collaboration~\cite{Grange:2015fou}
which was announced recently~\cite{Abi:2021gix},
is within $\fnalbnlsig\,\sig$ in agreement with  the older BNL result on \gmin2.
The combination of the two results 
yields a new deviation from the SM prediction of
\begin{align}
\De\amu^{\rm new} = (\newdiff \pm \newdiffunc) \times 10^{-10},
\label{gmtdiff-new}
\end{align}
corresponding to a discrepancy of $\newdiffsig\,\sig$.
The deviation in \refeq{gmtdiff-new} can easily be explained
in the realm of the Minimal Supersymmetric Standard Model  
(MSSM)~\cite{susyrefs} with
electroweak (EW) supersymmetric (SUSY) particle masses around
a few hundred~GeV. In these proceedings, following \citeres{CHS1,CHS3}, we
present an analysis of the parameter space of
MSSM that can accommodate the \gmin2\ result
while simultaneously being in agreement with the latest
direct search constraints from the LHC as well as the
constraints from dark matter (DM) relic density and direct detection (DD).
We assume that the lightest
SUSY particle (LSP), given by the lightest neutralino, $\neu1$,
makes up the full DM content of the
universe
\footnote{
In \citere{CHS2} we updated the analysis using the
DM relic density only as an upper bound.}.
We include the latest LHC searches 
via recasting in \CM~\cite{Drees:2013wra,Kim:2015wza, Dercks:2016npn}.
We find that the combined data helps to narrow down
the allowed parameter region, providing clear targets for possible future
colliders.

\section{The EW sector of MSSM}
\label{sec:model-constraints}

We give a very brief description of the EW sector of 
MSSM, consisting of
charginos, neutralinos and sleptons. 
The masses and mixings of the charginos and neutralinos are determined
by $U(1)_Y$ and $SU(2)_L$ gaugino masses $M_1$ and $M_2$, the Higgs
mixing parameter $\mu$ and the ratio of the two
vacuum expectation values (vevs) of the two Higgs doublets of MSSM,
$\tb = v_2/v_1$, all taken to be real.
This results in four neutralinos and two charginos 
with the mass ordering $\mneu1 < \mneu2 < \mneu3 <\mneu4$
and $\mcha1 < \mcha2$. Considering the size and sign of the anomaly,
we focus on positive values of $M_1$, $M_2$ and $\mu$
~\cite{CHS1}. 
For the sleptons, we choose common soft
SUSY-breaking parameters for all three generations, $\mL$ and $\mR$. 
We take the trilinear coupling
to be zero for all the three generations of
leptons ($l = e,\mu,\tau$). 
In general we follow the convention that $\Sl_1$ ($\Sl_2$) has the
large ``left-handed'' (``right-handed'') component.
The symbols are equal for all three generations, $\msl1$ and $\msl2$, but we
also refer to scalar muons directly, $\msmu1$ and $\msmu2$.

Following the stronger experimental limits from the
LHC~\cite{ATLAS-SUSY,CMS-SUSY},
we assume that the colored sector of the MSSM is decoupled from the EW sector.
We also assume that the stop masses in the TeV range
provide radiative corrections necessary to bring the light
$\cp$-even Higgs boson mass in the experimentally observed region,
$\Mh \sim 125 \gev$~\cite{Bagnaschi:2017tru,Slavich:2020zjv}.
$\MA$ has also been set to be above the TeV scale.


\section {Relevant constraints}
\label{sec:constraints}

The most important constraint that we consider comes from the \gmin2\
result. We use \refeq{gmtdiff-new} as a cut at the $\pm2\,\sig$ level.
We note that 
the main contribution to \gmin2\ in MSSM at the one-loop level comes from
diagrams involving $\cha1-\Sn$ and $\neu1-\tilde \mu$ loops. 
In our analysis the MSSM contribution to \gmin2\
at two loop order is calculated using {\tt GM2Calc}~\cite{Athron:2015rva},
implementing two-loop corrections
from \cite{vonWeitershausen:2010zr,Fargnoli:2013zia,Bach:2015doa}
(see also \cite{Heinemeyer:2003dq,Heinemeyer:2004yq}).

Various other constraints that are taken into account
comprises the following.
All points are checked to possess a stable and correct EW vacuum, e.g.\
avoiding charge and color breaking minima, using
the public code {\tt Evade}~\cite{Hollik:2018wrr,Ferreira:2019iqb}.
All relevant EW SUSY searches from the LHC are taken into account, mostly via
\CM~\cite{Drees:2013wra,Kim:2015wza, Dercks:2016npn}, where many
analyses had to be implemented newly~\cite{CHS1}.
For dark matter relic density
we use the latest result from Planck~\cite{Planck} :
$\Omega_{\rm CDM} h^2 \; = \; 0.120 \pm 0.001$.
We employ the constraint on the spin-independent
DM scattering cross-section $\ssi$ from
XENON1T~\cite{XENON}.


\section{Parameter scan}
\label{sec:scan}
\smallskip
The  scan regions  that cover the  
parameter space under consideration are as given below.
A more complete coverage of the MSSM parameter space can be found in \citeres{CHS1,CHS2,CHS3}.
\begin{description}
\item
{\bf (A) bino/wino DM with \boldmath{$\chapm1$}-coannihilation}
\begin{align}
  100 \gev \leq M_1 \leq 1 \tev \;,
  \quad M_1 \leq M_2 \leq 1.1 M_1\;, \notag \\
  \quad 1.1 M_1 \leq \mu \leq 10 M_1, \;
  \quad 5 \leq \tb \leq 60, \; \notag\\
  \quad 100 \gev \leq \mL \leq 1 \tev, \; 
  \quad \mR = \mL~.
\label{cha-coann}
\end{align}

\item
{{\bf (B) bino DM with \boldmath{$\Slpm$}-coannihilation}}: Case-R (SU(2) singlet)
\begin{align}
  100 \gev \leq M_1 \leq 1 \tev \;,
  \quad M_1 \leq M_2 \leq 10 M_1 \;, \notag\\
  \quad 1.1 M_1 \leq \mu \leq 10 M_1, \;
  \quad 5 \leq \tb \leq 60, \; \notag\\
  \quad M_1 \gev \leq \mR \leq 1.2 M_1, 
  \quad M_1 \leq \mL \leq 10 M_1~.
\label{slep-coann-singlet}
\end{align}
\end{description}

In all the scans we choose flat priors of the parameter space and
generate \order{10^7} points. A detailed account of our numerical set up
and analysis flow can be found in \citere{CHS1}.


\section{Results}
\label{sec:results}
\begin{figure}[htb!]
\begin{subfigure}[b]{0.48\linewidth}
        \centering\includegraphics[width=0.8\textwidth]{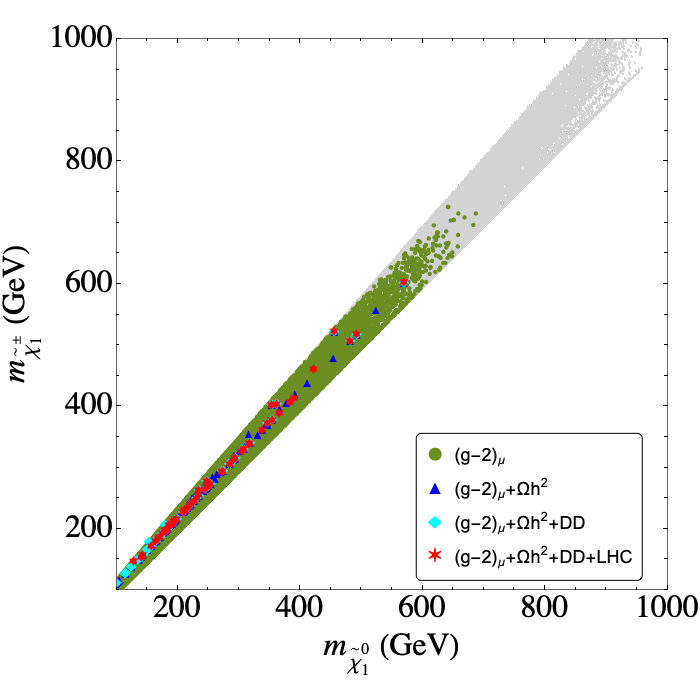}
        \caption{}
        \label{}
\end{subfigure}
~
\begin{subfigure}[b]{0.48\linewidth}
        \centering\includegraphics[width=0.8\textwidth]{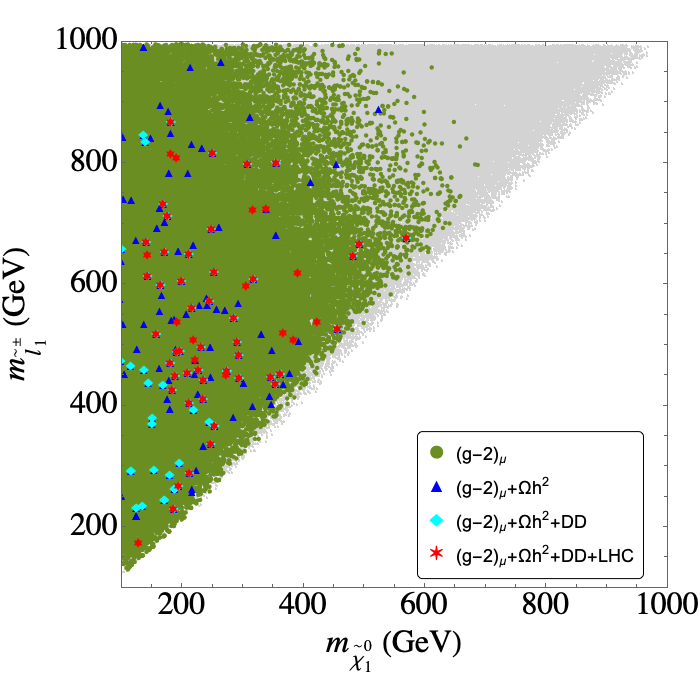}
        \caption{}
        \label{}
\end{subfigure}
\caption{The results of our parameter scan for  the bino/wino
        $\cha1$-coannihilation scenario in the $\mneu1$--$\mcha1$ plane
        (left) and $\mneu1$--$\msl1$ plane (right). 
For the color coding: see text.
}
\label{fig:charco}
\end{figure}

In this section we review some of the results for the 
scenarios defined above~\cite{CHS1,CHS3}. 
We  denote the points surviving certain constraints
with different colors. In grey (round) we show all of our scan points.
In green (round), blue (triangle), cyan (diamond) and red (star) we show points that
additionally pass the \gmin2, correct relic density, DD 
and the LHC constraints respectively.
We start with the results in the
$\cha1$-coannihilation scenario in \reffi{fig:charco}. 
In the $\mneu1$--$\mcha1$ plane, shown in the left plot, by definition
of $\cha1$-coannihilation the points are clustered along the diagonal of
the plane. One observes a clear upper limit on the (green) points
allowed by the new \gmin2\ result of about $\htrs{700} \gev$. 
Applying the CDM constraints reduce the upper limit further to
about $\htrs{600} \gev$.
Applying the LHC constraints, corresponding to the ``surviving'' red points (stars), does
not yield a further reduction from above, but cuts always 
only points in the lower mass range.
Thus, the experimental data set an upper as well as a lower bound,
yielding a clear search target for the upcoming LHC runs as well as
for future electron-positron colliders.
The distribution of the lighter slepton mass (where it should be kept in
mind that we have chosen the same masses for all three generations)
is presented in the $\mneu1$-$\msl1$ plane, shown
in the right plot of \reffi{fig:charco}. The LHC constraints
which are most important in this scenario comes from  slepton pair
production leading to two leptons and missing energy in the final state~\cite{Aad:2019vnb}
and  compressed spectra searches~\cite{Aad:2019qnd}.

\begin{figure}[htb!]
\centering
\begin{subfigure}[b]{0.48\linewidth}
        \centering\includegraphics[width=0.8\textwidth]{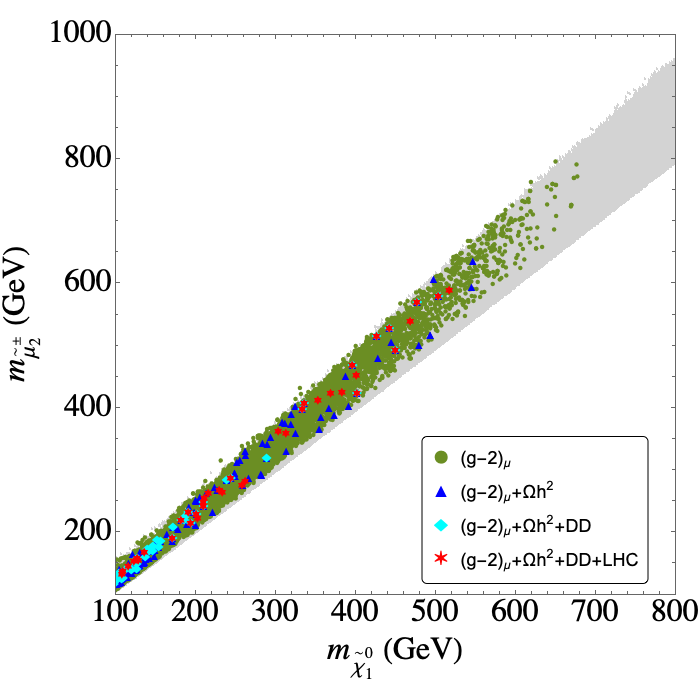}
        \caption{}
        \label{}
\end{subfigure}
~
\begin{subfigure}[b]{0.48\linewidth}
        \centering\includegraphics[width=0.8\textwidth]{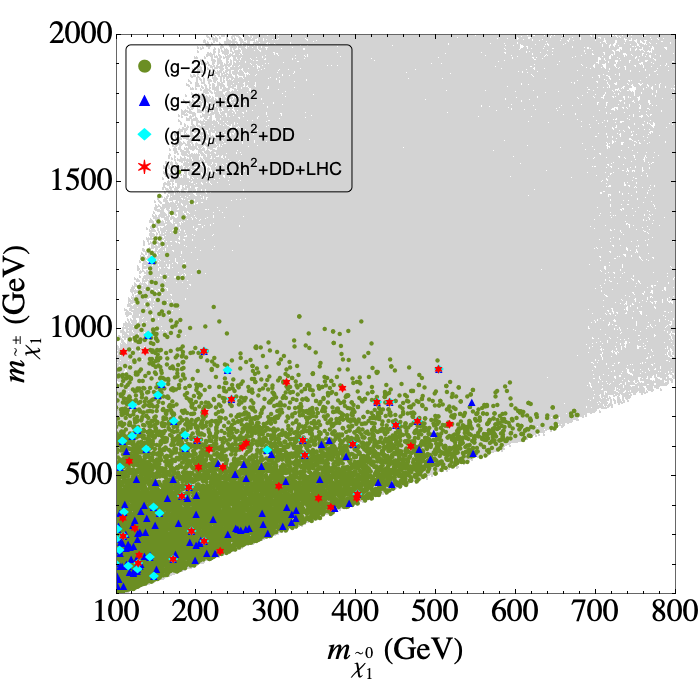}
        \caption{}
        \label{}
\end{subfigure}
\caption{The results of our parameter scan for  the 
        $\Slpm$-coannihilation case-R scenario in the $\mneu1$--$\msmu2$ plane 
        (left) and $\mneu1$--$\mcha1$ plane (right). 
For the color coding: see text.
}
\label{fig:caseR}
\end{figure}

We now turn to the scenario of $\Slpm$-coannihilation Case-R,
where in the scan we require the ``right-handed'' sleptons to be close
in mass to the LSP. It should be kept in mind that in our notation
the ``left-handed'' (``right-handed'')
slepton corresponds to $\Sl_1$ ($\Sl_2$).
We start in \reffi{fig:caseR} with the $\mneu1$-$\msmu2$ plane
in the left plot. By definition
of the scenario the points are concentrated along the diagonal. The 
\gmin2\ bound yields an upper limit on the LSP mass of
$\sim \htrs{690} \gev$. The \gmin2\ bound also places
an upper limit on $\msmu2$ (which is close
in mass to the $\Sel2$ and $\Stau2$) of $\sim \htrs{800} \gev$.
Including the CDM and LHC constraints, these limits reduce to
$\sim \htrs{520} \gev$ for the LSP, 
and correspondingly to $\sim \htrs{600} \gev$ for $\msmu2$
and $\sim \htrs{530} \gev$ for $\mstau2$.
In the right plot of \reffi{fig:caseR} we show the results in the
$\mneu1$-$\mcha1$ plane. The upper limits on the
chargino mass are
reached at $\sim \htrs{900} \gev$, including all the constraints. 
The most relevant LHC constraints
in this scenario comes from  $\chapm1-\neu2$ pair
production leading to three leptons and missing energy in the final state~\cite{Aaboud:2018jiw},
direct slepton pair production searches~\cite{Aad:2019vnb}
and  compressed spectra searches~\cite{Aad:2019qnd}.

\medskip

\section*{Acknowledgments}
\small
I.S.\ thanks S.~Matsumoto for the cluster facility.
The work of I.S.\ is supported by World Premier
International Research Center Initiative (WPI), MEXT, Japan.
The work of S.H.\ is supported in part by the
MEINCOP Spain under contract PID2019-110058GB-C21 and in part by
the AEI through the grant IFT Centro de Excelencia Severo Ochoa SEV-2016-0597.
The work of M.C.\ is supported by the project AstroCeNT:
Particle Astrophysics Science and Technology Centre,  carried out within
the International Research Agendas programme of
the Foundation for Polish Science financed by the
European Union under the European Regional Development Fund.


\newcommand\jnl[1]{\textit{\frenchspacing #1}}
\newcommand\vol[1]{\textbf{#1}}

\newpage{\pagestyle{empty}\cleardoublepage}


\end{document}